\documentstyle[multicol,aps,floats,epsfig]{revtex} 

\begin{document}

\draft
\twocolumn[\hsize\textwidth\columnwidth\hsize\csname@twocolumnfalse\endcsname
\title{Dephasing and Metal-Insulator Transition}
\author{Junren Shi and X. C. Xie} 
\address{ Department of Physics, Oklahoma State University,
  Stillwater, OK 74078 }
\maketitle
\begin{abstract}
  The metal-insulator transition (MIT)
  observed in two-dimensional (2D) systems
  is apparently contradictory to the well known scaling theory of
  localization. By investigating the conductance of disordered
  one-dimensional systems with a finite phase coherence length, we show
  that by changing the phase coherence length or the localization
  length, it is possible to observe the transition from insulator-like
  behavior to metal-like behavior, and the transition is a
  crossover between the quantum and classical regimes.  The resemblance
  between our calculated results and the experimental 
  findings of 2D MIT suggests that the
  observed metallic phase could be the result of a finite dephasing rate.
\end{abstract}
\bigskip]
Since the discovery of the metal-insulator transition in
two-dimensional (2D) systems\cite{Kravchenko}, several theoretical
models have been proposed to understand the phenomena. Among the
proposed theories, some\cite{Xie,Sarma,Altshuler,Meir} can be
considered as semi-classical theory. Although they are different in
details, the basic idea is the same.  They all consider that the
metallic phase observed in the experiments is the classical phase,
\emph{i.e.,} the metallic behavior of the system can be well
understood under the classical picture.  For instance, the percolation
model, initially proposed by He and Xie\cite{Xie} and further extended
by Meir\cite{Meir}, provides good description of many experimental
facts. In this approach, the system consists of inhomogeneous carrier
distribution with high density conducting regions and low density
insulating regions. By assuming that the electrons are totally
dephased in each separate region, one can consider the metal-insulator
transition as the classical percolation transition of the high density
conducting regions, and calculate the total conductance of the system
by classical random resistance network. The dephasing of the carriers
is essential for the model since a pure quantum system will never
percolate according to the well-known scaling theory of
localization\cite{Anderson}.

It is not obvious that a system can be considered as a classical
system at low temperatures.  A recent experiment \cite{Pudalov} on 2D
systems shows that the phase coherence length is quite long, typically
600-1000 nm.  Thus, it is more likely that a real system is in the
regime where quantum effects compete with classical effects. Classical
effects are manifested by a finite phase coherence length, which may
be due to a finite temperature or other novel mechanisms. For
instance, there are experimental indications that the dephasing rate
may be finite even at zero temperature \cite{Mohanty}.  On the other
hand, in a disordered 2D system, the quantum effect always causes the
localization length to be finite.  For a 2D system with
metal-insulator transition, the localization length strongly depends
on the carrier density. Actually, by changing the carrier density, the
conductance of the system may change several orders \cite{Kravchenko}
which implies a substantial change in the localization length.
The behavior of the system is determined by two competing length
scales: localization length and phase coherence length.  Therefore, it
is important to study transport properties by varying these two
lengths.

\begin{figure}
  \centering
  \epsfig{file=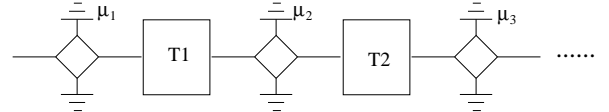, width=0.9\columnwidth}
  \caption{The model to simulate the system with dephasing and disorder.
    The rectangles represent the normal random scatters, and the
    diamonds represent the dephasing scatters.}
  \label{fig: 1}
\end{figure}

In this paper, we study the interplay of disorder and dephasing of
one-dimensional (1D) systems in transport properties.  We limit to 1D
models to reduce the severe finite-size effect in numerical results
for higher dimensions, although our conclusion can be carried over to
higher dimensions. Our 1D model (Fig.1) consists of normal random
scatters and dephasing scatters, alternatively. While the normal
random scatters give rise to a finite localization length, the
dephasing scatters randomize electron phase.  The normal random
scatters are constructed by \( M \) \( \delta \)-barriers with random
height \( q_{i} \), which has the distribution
\[
P(q_{i})=\left\{ \begin{array}{cl}
    \frac{1}{\Delta q}, & {\rm if}\, 
    -\frac{\Delta q}{2}\leq q_{i}\leq \frac{\Delta q}{2}\\
    0, & {\rm otherwise}
\end{array}\right. .\]
In the model $\Delta q$ controls the randomness of the system.  The
transmission and reflection coefficients for the normal random
scatters can be calculated from the transfer-matrix for individual \(
\delta \)-barrier \cite{Abrahams},
\[
U_{i}=\left[ \begin{array}{cc}
    \frac{1}{t^{*}} & \frac{r}{t}\\
    \frac{r^{*}}{t^{*}} & \frac{1}{t}
\end{array}\right] =\left[ \begin{array}{cc}
1-i\frac{q_{i}}{2k} & -i\frac{q_{i}}{2k}\\
i\frac{q_{i}}{2k} & 1+i\frac{q_{i}}{2k}
\end{array}\right], \]
where \( t \) and \( r \) are the transmission and reflection
amplitudes for the barrier, and \( k \) is the momentum of the
injected electron. The transfer matrix for \( M \) sequential \(
\delta \)-barriers can be calculated from

\[
U^{M}=U_{M}XU_{M-1}X\cdots U_{1}X,\] where \( X \) is the transfer
matrix describing the propagation of the electron from one \( \delta
\)-barrier to the next. Assuming the spacing between the neighboring
barriers is unity, $X$ is
\[
X=\left[ \begin{array}{cc}
    e^{ik} & 0\\
    0 & e^{-ik}
\end{array}\right] .\]
Using the transfer-matrix technique, the localization length can be
determined analytically \cite{Abrahams}.

\begin{figure}
  \centering
  \epsfig{file=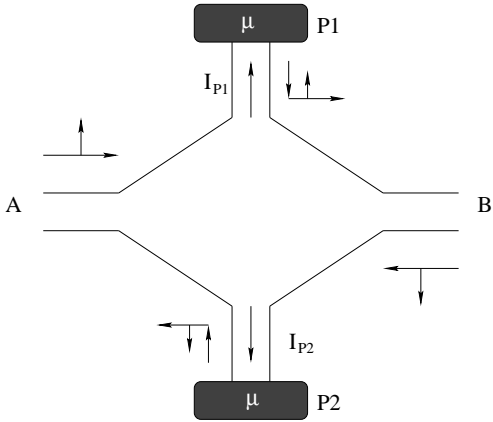, width=0.75\columnwidth}
  \caption{The structure of an individual dephasing scatter. 
    The system is connected to two identical external electron
    reservoirs P1 and P2.  An electron can be scattered along the
    directions shown by the arrows. There is no backscattering caused
    by this dephasing scatter.}
  \label{fig: 2}
\end{figure}

To introduce the dephasing effect into the system, one has to include
the interaction between the system and the environment. The
B\"{u}ttiker model \cite{Buttiker} shown in Fig.\ref{fig: 2} is the
simplest way to achieve that.  In this approach, the system is
connected to the external electron reservoirs via the dephasing
scatters. With a certain possibility, an electron is scattered into
the external reservoirs, totally losing its phase memory, and then
re-injected into the system.  Two restrictions are imposed to reflect
physical reality.  First, the net current between the system and the
reservoirs should be zero so that each scattered electron will finally
return to the system. To do so, one can adjust the chemical potential
\( \mu \) of the external reservoirs such that \( I_{P1}+I_{P2}=0, \)
where \( I_{P1} \)(\( I_{P2} \)) is the current between the system and
the external reservoir P1 (P2). Second, the system is connected to two
identical electron reservoirs P1 and P2, and the \( S \)-matrix
between the system and the reservoirs is designed so that the electron
is only scattered forward, thus the dephasing scatters will not cause
any momentum relaxation.  The \( S \)-matrix reads \cite{Buttiker}
\[
S=\begin{array}{cc}
\begin{array}{c}
A\\
B\\
P1\\
P2
\end{array} & \left[ \begin{array}{cccc}
0 & \sqrt{1-\alpha } & 0 & -\sqrt{\alpha }\\
\sqrt{1-\alpha } & 0 & -\sqrt{\alpha } & 0\\
\sqrt{\alpha } & 0 & \sqrt{1-\alpha } & 0\\
0 & \sqrt{\alpha } & 0 & \sqrt{1-\alpha }
\end{array}\right] 
\end{array},\]
where \( \alpha \) is the possibility that the electron is scattered
to the reservoirs, namely, the dephasing rate. The phase coherence
length is estimated by \( L_{\varphi }\simeq M/\alpha \).

The localization length of the system is determined by the normal
random scatters, while the phase coherence length is determined by the
dephasing scatters. For a system with \( N \) dephasing scatters,
there are \( N+2 \) external chemical potentials \( \mu _{i} \) with
\( i=L,\, R,\, 1,\, 2,\, \cdots N \), where \( \mu _{L(R)} \) is the
chemical potential for the left (right) measurement electrode. The
system satisfies the multi-lead Ohm's law,
\[
I_{i}=\sum _{j}\sigma _{ij}\mu _{j},\, \, \, \, \, \, \, i,\, j=L,\,
R,\, 1,\, 2,\, \cdots N.\] The conductance between the leads, \(
\sigma _{ij} \), is determined by the Landaur-B\"{u}ttiker formula,
\[
\sigma _{ij}=\frac{2e^{2}}{h}T_{ij}\, \, \, \, \, \, \, {\rm for}\,
i\neq j,\] where \( T_{ij} \) is the transmission coefficient between
the leads. For the leads attached to the dephasing scatters, the
transmission coefficient is the sum of all transmission coefficients
between leads P1 and P2,
\begin{eqnarray*}
T_{ij} & = & T_{i\, P1,j\, P1}+T_{i\, P1,j\, P2}+T_{i\, P2,j\, P1}
+T_{i\, P2,j\, P2},\\
 &  & {\rm for}\, i,j=1,\, 2,\, \cdots N.
\end{eqnarray*}
The gauge invariance, namely shifting each $\mu _{i}$ by a constant
should not affect the result, is satisfied through the condition
\[
\sigma _{ii}=-\sum _{j\neq i}\sigma _{ij}.
\] 
The total conductance of the system is calculated by imposing the
condition
\[
I_{i}=0\, \, \, \, \, {\rm for}\, i=1,\, 2,\, \cdots N.
\]
After some algebra, the total conductance of the system can be written
as
\[
\sigma _{tot}=\sigma _{LR}-\left[ \begin{array}{cccc} \sigma _{L1} &
    \cdots & \sigma _{LN}
\end{array}\right] \left[ \begin{array}{ccc}
\sigma _{11} & \cdots  & \sigma _{1N}\\
\vdots  & \ddots  & \vdots \\
\sigma _{N1} & \cdots  & \sigma _{NN}
\end{array}\right] ^{-1}\left[ \begin{array}{c}
\sigma _{1R}\\
\vdots \\
\sigma _{NR}
\end{array}\right] .\]
Although we start from a seemly artificial model, the total
conductance formula actually reflects physical reality.  It contains
two kinds of contributions: the first is the direct conductance \(
\sigma _{LR} \) coming from direct quantum tunneling; the second is
the correction due to the dephasing effect caused by classical
sequential tunneling. Thus, the conductance formula is consistent with
the general picture of the dephasing effect on a conductance.

\begin{figure}
  \centering 
  \epsfig{file=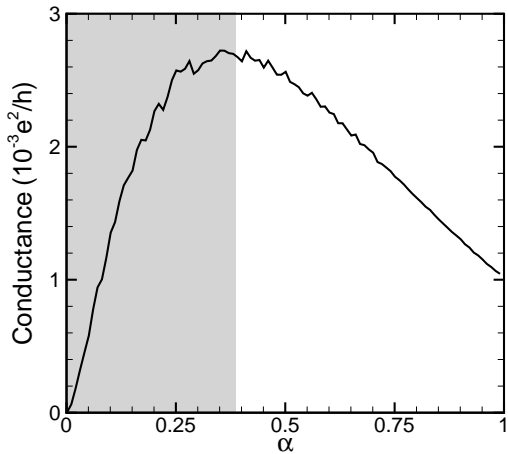, width=0.9\columnwidth}
  \caption{The typical behavior of the conductance. We use the 
    parameters \protect\( Q_{0}=\overline{q/2k}=1.1\protect \),
    \protect\( \Delta q=0.5\protect \), \protect\( M=5\protect \),
    \protect\( N=100\protect \), and average over 500 samples. The
    gray and white regions show the quantum and classical regime
    respectively.}
  \label{fig: 3} 
\end{figure}

The typical behavior of the conductance for the system is shown in
Fig.\ref{fig: 3}, where we plot the conductance as a function of the
dephasing rate \( \alpha \).  In a real system, the dephasing rate is
a monotonic function of temperature, so the plot can also be
considered as the temperature dependence of the conductance. We have
systematically calculated the conductance for different sets of
parameters, and the results show qualitatively similar behavior,
although the peak position depends on the parameters. The most
important feature of the plot is that the conductance is not a
monotonic function of the dephasing rate, or temperature.  The plot
can be divided into two regions, the gray and white regions.  In the
gray region, the dephasing rate is low and the phase coherence length
is long.  The electron is localized within the phase coherence length,
so the system shows quantum localization with insulator-like behavior,
namely, the conductance decreases as temperature drops.  This region
can be considered as the quantum region.  On the other hand, when the
dephasing rate becomes higher, the system enters into the classical
regime and the conductance increases with decreasing temperature, a
typical metallic behavior.

\begin{figure}
  \centering
  \epsfig{file=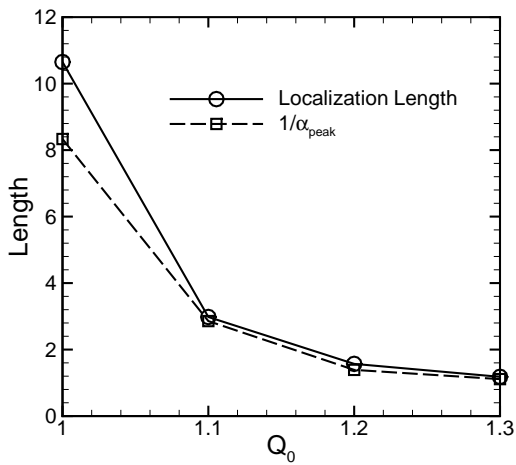, width=0.9\columnwidth}
  \caption{The comparison between the localization length and the 
    \protect\( 1/\alpha _{peak}\protect \), which is equivalent to the
    phase coherence length. \protect\(
    Q_{0}=\overline{q_{i}/2k}\protect \), and \protect\( \Delta
    q=0.5\protect \). }
  \label{fig: 4} 
\end{figure}

The turning point between the quantum and classical regimes can be
determined by comparing the phase coherence length \( L_{\varphi } \)
with the localization length \( \xi \) calculated from the transfer
matrix formalism \cite{Abrahams}.  The result is shown in
Fig.\ref{fig: 4}.  The phase coherence length \( L_{\varphi} \)
is obtained by using the value of $\alpha$ corresponding to
the peak in the conductance
plot (see Fig.3) for a given $Q_{0}$. One can see that the two lengths
are approximately equal. Thus, it shows
that the transition occurs at the
point where \( L_{\varphi }\sim \xi \).  From Fig.\ref{fig: 3} and
Fig.\ref{fig: 4}, one can clearly see that by changing the phase
coherence length or the localization length, it is possible to observe
the transition from the metal-like behavior to the insulator-like
behavior, and the behavior of the system is determined by the
competition between the localization length and the phase coherence
length.  When \( L_{\varphi }\sim \xi \), the quantum physics of
localizaiton ceases to exist.
Therefore, we believe that it is the transition between the quantum and the
classical phases.

\begin{figure}
  \centering
  \epsfig{file=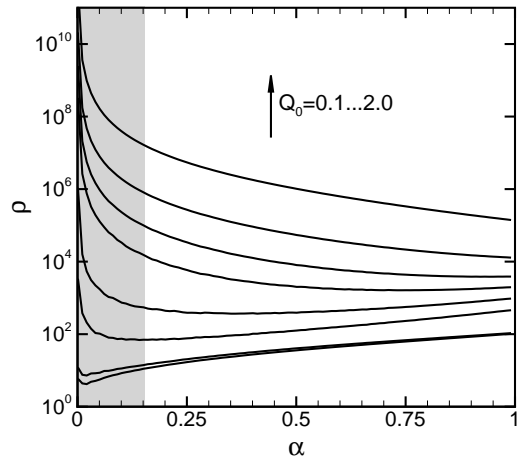, width=0.9\columnwidth}
  \caption{The dephasing rate (or temperature)
    dependence of the resistance for different carrier densities
    (\protect\( Q_{0}=\overline{q_{i}/2k}\protect \)). The gray region
    shows inaccessible region due to the finite cutoff of the
    dephasing rate. We use \protect\( \Delta q=0.5\protect \), and
    \protect\( Q_{0}=\protect \)0.1, 0.5, 1.0, 1.1, 1.2, 1.3, 1.5,
    2.0.}
  \label{fig: 5}
\end{figure}
\begin{figure}
  \centering
  \epsfig{file=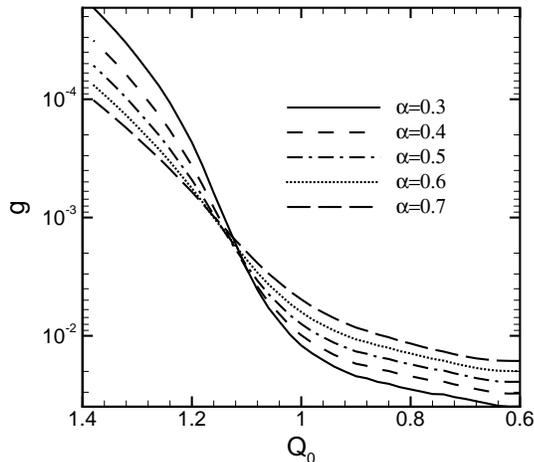, width=0.9\columnwidth}
  \caption{The density dependence of the conductance for different 
    temperatures. We reverse the direction of the axes so that it is
    easier to compare with \protect\( n-\rho \protect \) plot wildly
    used in the literatures.}
  \label{fig: 6}
\end{figure}

Great resemblance can be found when comparing the experimental
findings\cite{Kravchenko} with the results from this simplified model.  In
Fig.\ref{fig: 5} we plot the dephasing rate (or temperature)
dependence of the resistance. Different curves are for different
scattering parameter $Q=q/2k$. Changing the carrier density is
equivalent to changing the Fermi momentum, which in turn changes $Q$.
Thus, different curves correspond to different carrier densities. In
the plot, we impose a finite cutoff of the dephasing rate $\alpha$,
which makes lower $\alpha$ inaccessible (gray region in the plot). For
$\alpha$ above the cutoff, the phase coherence length $L_{\phi}$ is
always finite. Depending on whether $L_{\phi}$ is larger or smaller
than the localization length $\xi$, one can either observe the
metal-like or the insulator-like behavior, as shown in Fig.\ref{fig:
  5}. When the finite cutoff falls upon the turning point (the maximum
conductance point in Fig.3), the system shows the ``critical''
behavior, where the resistance is nearly flat within a certain
temperature range.  In Fig.\ref{fig: 6}, we show the density
dependence of the conductance for different $\alpha$ (or temperature).
One can identify a ``fixed'' point where different curves cross at.
Similar feature has been seen in many experimental plots.

The finite cutoff of the dephasing rate can be justified by two
possible reasons.  First, the cutoff may be due to a finite
temperature. In this case, if we assume the dephasing rate goes to
zero when temperature approaches zero, as suggested by a simple power
law \( \alpha \sim T^{\nu } \), there is always an upturn of the
resistance at low enough temperature as shown in Fig.5 for low values
of $\alpha$.  This suggests a re-entrance to an insulator at low
temperatures.  The re-entry behavior may have already been observed in
the recent experiment\cite{Simmons}. Under the circumstance, the
transition is a finite temperature effect.  The second possibility is
that the dephasing rate might be finite at zero
temperature\cite{Mohanty}.  Consequently, the metallic phase will
survive even at zero temperature. If we adhere to the original
definition that a metal has a finite resistance at zero temperature
while the resistance of an insulator diverges, the system will always
be a metal. The reason is that on the low density "insulator" side,
the resistance will saturate to a finite value at $T=0$ because of the
finite phase coherence length. However, in a similar plot as shown in
Fig.6, a "fixed" point can still be identified which can be used as an
operational definition of "metal-insulator transition".

The saturation of the dephasing rate at low temperatures is still a
controversial issue. Some argue that the saturation observed in the
experiments is not an intrinsic effect.  Nevertheless, whether it is
intrinsic or extrinsic, the same factors which cause the saturation
should have a similar effect on the conductance.  To justify a real
metal-insulator transition, one has to clearly rule out those external
factors that may cause finite dephasing rate at low temperatures
\cite{Mohanty}.


In summary, we have studied the interplay between dephasing and
disorder. Based on a 1D model, we show that by changing the phase
coherence length or the localization length, it is possible to observe
the transition from the insulator-like behavior to the metal-like
behavior, which corresponds to a transition between quantum and
classical phases.  The great resemblance between the results
from this simplified model
and the experiments suggests that the quantum effect is important at
low temperatures, although the high temperature behavior is dominated
by the classical effect.  We suggest that conductance experiment
should be accompanied by a dephasing rate measurement to address the
effect of a finite coherence length. Although our calculation is for a
1D model, the same physics should survive at higher dimensions.

The work is supported by DOE.


\begin{thebibliography}{}
\bibitem{Kravchenko}S. V. Kravchenko \emph{et. al.}, Phys. Rev. B
  \textbf{51}, 7038 (1995); S. V.  Kravchenko \emph{et. al.}, Phys.
  Rev. Lett. \textbf{77}, 4938 (1996).
%
\bibitem{Xie}Song He and X. C.  Xie, Phys. Rev. Lett. \textbf{80},
  3324 (1998); Junren Shi, Song He and X. C. Xie, Phys. Rev. B
  \textbf{60}, \textbf{R}13950 (1999); Junren Shi, Song He and X. C.
  Xie, cond-mat/9904393.
%
\bibitem{Sarma}S. Das Sarma and E. H. Hwang, Phys. Rev. Lett.
  \textbf{83}, 164 (1999).
%
\bibitem{Altshuler}B. L. Altshuler and D.  L. Malsov, Phys. Rev. Lett.
  \textbf{82}, 145 (1999); B. L. Altshuler and D. L. Malsov, Phys.
  Rev. Lett. \textbf{83}, 2092 (1999).
%
\bibitem{Meir}Y. Meir, Phys.  Rev. Lett. \textbf{83}, 3506 (1999).
%
\bibitem{Anderson}E. Abrahams, P. W. Anderson, D. C. Licciardello and
  T. V. Ramakrishnan, Phys.  Rev. Lett. \textbf{42}, 673 (1979).
%
\bibitem{Pudalov}G. Brunthaler, A. Prinz, G. Bauer, V. M. Pudalov, E.
  M. Dizhur, J. Jaroszynski, P.Gold and T. Dietl, preprint
  cond-mat/9911011.
%
\bibitem{Mohanty}P.  Mohanty, E. M. Q. Jariwala, R. A. Webb, Phys.
  Rev. Lett.  \textbf{78}, 3366 (1997); P. Mohanty and R. A. Webb,
  Phys. Rev. B \textbf{55}, R13452 (1997).
%
\bibitem{Abrahams}B. S. Andereck, and E. Abrahams, \textit{Physics in
    One Dimension}, edited by J.  Bernasconi and T. Schneider
  (Springer-Verlag, 1981).
%
\bibitem{Buttiker}M. B\"{u}ttiker, Phys.  Rev. B \textbf{33}, 3020
  (1986) and references therein; S. Datta, \textit{Electronic
    Transport in Mesoscopic Systems} (Cambridge University Press,
  1995).
%
\bibitem{Simmons}M. Y. Simmons, A. R.  Hamilton, M. Pepper, E. H.
  Linfield, P. D. Rose and D.  A. Ritchie, Phys. Rev. Lett.
  \textbf{84}, 2489 (2000).
\end{thebibliography}
\end{document}